# Building and Operating A Large Scale Storage Facility

## Brookhaven National Laboratory's RHIC Computing Facility

Maurice Askinazi, David Free,
Bruce Gibbard, Thomas Throwe

When considering past and current successes in the building and operation of the RHIC Computing facility's General Computing Environment three principles have proven helpful in developing a useful work environment for our scientists while also keeping our facility manageable.

- *Consolidation*
- *Documentation*
- *High Availability*

### Consolidation

refers to both Hardware and Software platforms. By having a single hardware platform such as SUN servers, we are able to maintain a small spare parts cache that is very effective as well as have expertise in dealing with common hardware issues.
By having a single Operating System to maintain we are able to maintain a high level of expertise in it's management, have disk images that reflect various patch levels and OS releases for our systems, provide a software repository for programs built on this platform, as well as stay on top of security and bug fixes for this operating environment.

Below I've presented three evolutions of our server hardware environment. We began with system cabinets containing a mix of hardware. Having a variety of vendors caused problems with purchasing and support. Various manufacturers had different methods of operation and it was difficult to be efficient in planning consumption of space and power.

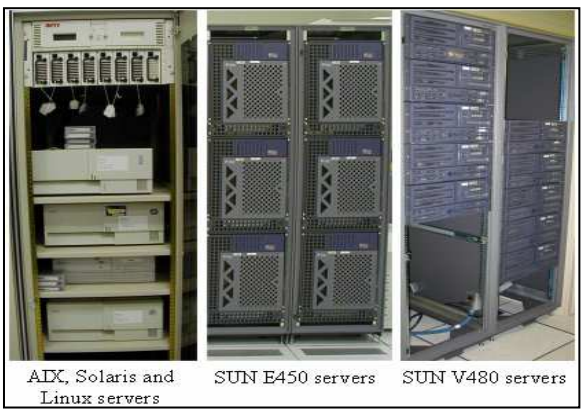

Over time, we've migrated to a system of racks containing servers of similar manufacturer, similar operating system.

Below I've presented several evolutions of our disk hardware environment. We began with direct connect RAID controllers. We grew by building a large Fibre Channel SAN and attaching high bandwidth, high availability, high capacity storage to it. Note the increase of density with each purchase.

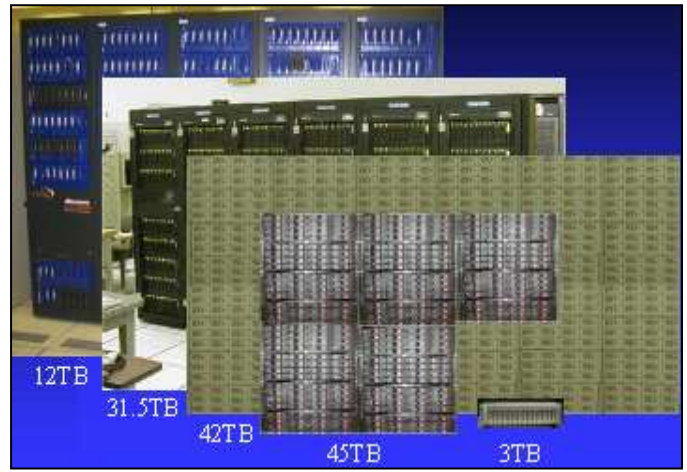

### Documentation

is necessary for several reasons, the most important ones being the tracking of what equipment you have in production, facilitating repair during emergencies, mapping ones current environment for the benefit of vendors working on future procurements, as well as the increased understanding of ones environment and skills which comes about from the process of trying to document it. Below I've provided screenshots of some of our more important documentation.

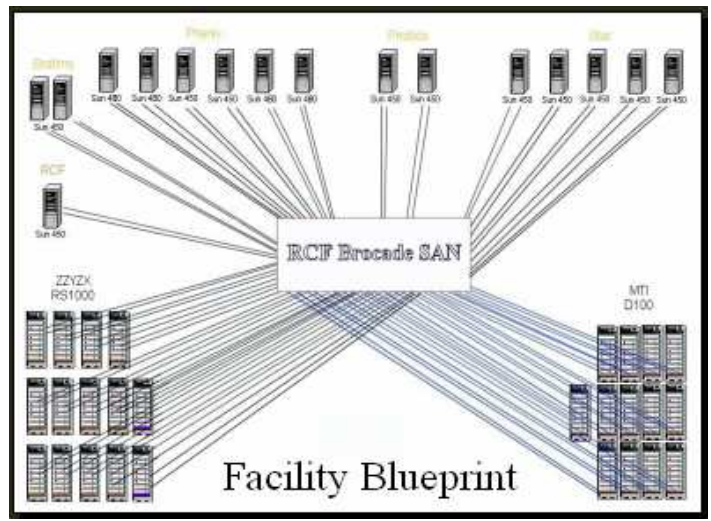

The "Facility Blueprint" is very helpful as a starting point for discussions with vendors and support technicians in conceptualizing our facility.

Each Fibre Channel device on a SAN (Storage Area Network) has a unique ID. We track every port on our SAN and what is plugged into it. Below is part of our list of devices associated with their switch ports and unique World Wide Port ID.

| System Name | Domain # | Port# | Port_WWN |
|---|---|---|---|
| MTI00 | 158 | 0 | 10:00:00:00:c9:22:2b:a9 |
| MTI00 | 157 | 0 | 10:00:00:00:c9:23:04:cf |
| MTI01A | 130 | 2 | 21:00:0c:c0:01:00:16:bd |
| MTI01B | 135 | 2 | 23:00:0c:c0:01:00:16:bd |
| MTI02A | 131 | 4 | 21:00:0c:c0:01:00:05:27 |
| MTI02B | 134 | 4 | 23:00:0c:c0:01:00:05:27 |
| MTI03A | 132 | 1 | 21:00:0c:c0:00:00:05:cf |
| MTI03B | 133 | 1 | 23:00:0c:c0:00:00:05:cf |
| MTI04A | 132 | 0 | 21:00:0c:c0:01:00:09:15 |

Below is a segment of our page which lists which filesystems are mounted and served by which servers.

| Brahms | | |
|---|---|---|
| RMINE001 | RMINE002 | RMINE003 |
| /brahms/u<br><br>www<br><br>data07<br>data08<br>data09 | data01<br>data02<br>data03<br>data04<br>data05<br>data06 | data10<br>data11<br>data12<br>data13<br>data14 |

Below is a segment of our page which lists storage devices and which server they are mounted upon.

| Storage | Controller | LUN # | Server |
|---|---|---|---|
| ZZYZX01 | 131,132 | 0 | RMINE603 |
| | | 2 | RMINE603 |
| | 133,134 | 1 | RMINE603 |
| | | 3 | RMINE603 |
| ZZYZX02 | 135,136 | 0 | RMINE605 |
| | | 2 | RMINE605 |
| | 137,138 | 1 | RMINE605 |
| | | 3 | RMINE605 |
| ZZYZX03 | 139,140 | 0 | RMINE202 |
| | | 2 | RMINE202 |
| | 141,142 | 1 | RMINE202 |
| | | 3 | Available |

**High Availability**
is an essential concern when building a large facility because you cannot plan forward when you are tied up in issues regarding your current facility. Once equipment is installed it is the hope of both the user and system support that it will be stable and provide many years of reliable use.

High availability can be addressed on both the clients and servers.

**- Clients -**
should not be using NFS hard mounts to the server. Nor should they be using the server's true system name.

By using an automount daemon to mount the Network File System when it is needed, many NFS filesystem outages are missed by clients that are not currently using the NFS mount. They also resolve themselves when the resource returns. This allows most quick reboots of the servers to pass unnoticed by non-interactive batch programs running on the clients.

By using DNS aliases in your client's system files rather than true server names you are able to move storage to another server by making a single change in one place, the DNS database, rather than editing the system files on every possible client which may mount the exported filesystem.

DNS aliases are helpful in two roles. Virtualizing the server name and distributing load onto multiple machines.

The following is an example of a DNS lookup of a system alias:

# nslookup starnfs
Name: rnfs09.rcf.bnl.gov
Address: 13.19.26.4
Aliases: starnfs.rcf.bnl.gov

The following is an example of round-robin load sharing using DNS:

# nslookup rssh
Name: rssh.rcf.bnl.gov
Addresses: 13.19.6.7, 13.19.6.3, 13.19.6.4

### - Servers -
need to assess **"Single Point Of Failure"**

Items to be considered are:
- HBA (Host Bus adapter)
- Cabling
- Disk Controllers
- Disks

*Sometimes responding to SPOF introduces additional complexity and another new SPOF*

### HBA
When connecting external disks with a HBA, it is advantageous to have two adapters in the server to eliminate SPF.
This requires Multipathing software and a multiported disk system.

### Cabling
Multiple HBA / Disk Controller paths requires multiple cable connections.
Introduction of a switch in the cable path allows for greater access possibilities, higher availability, but introduces another SPOF.

### Disk Controllers
Their failover ability will determine if component failure is transparent to server, and if repair can be done online.
Also determines how component disks are configured for performance, reliability, or maximum capacity.

### Disks
Disks should be configured with some element of redundancy RAID 1,3,5 so that a single disk failure wont take the LUN offline.
System should provide for configuration of a hot spare, and possess the ability to perform automatic rebuild.

### Fibre Channel Switches
Implementing a single switch solution introduces a SPOF, but benefits are increased number of data paths.
There are suggested switch topologies which will remove SPOF. Recommended topology has a maximum of one hop to any other switch. Core / Edge Topology is utilized at RCF. Double Core switches, and a split fabric is highly resilient.

## Is High Availability Achieved?

We removed these SPOF:

- Root disk on server
- Server Power
- Server HBAs
- Cabling
- FC Switch
- FC Switch Power
- RAID power supplies
- Disk Controllers
- Disks

## High Availability Visualized

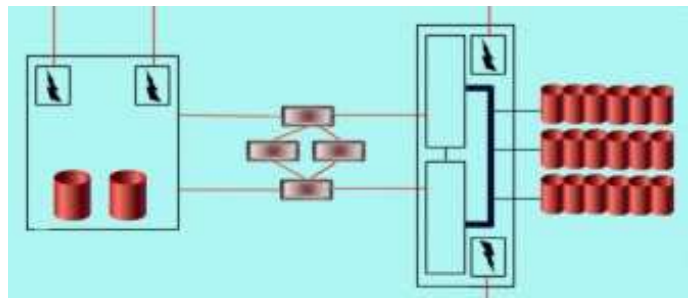

Server                Raid Controller

### Configuring Storage
as a vehicle to present component storage concepts.

- Map Storage to Server (SAN Zoning)
- Map LUN to Server (Disk Controller)
- Map Storage to Target (HBA Config)
- Mount Disk
- Import Disk to Veritas

Log on to SAN.
Tool used here is ***Brocade Fabric Manager***

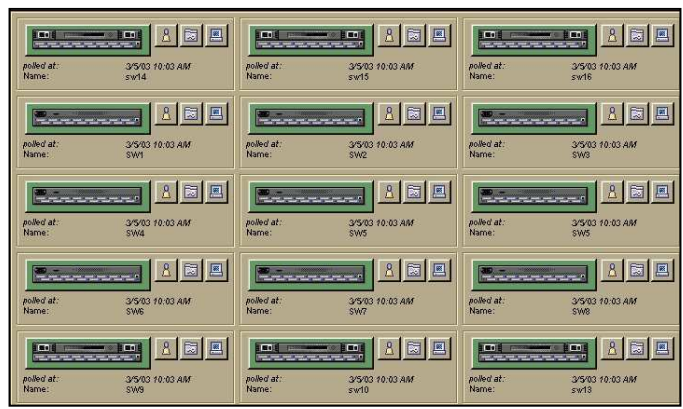

Open a "Name Server" Window to identify the "WWN device ID" of your Fibre Channel device by looking up the port that it is physically plugged in to.

| Name Server Table | | | | | | |
|---|---|---|---|---|---|---|
| Domain # | Port # | Port ID | Port Type | Port WWN | Node WWN | Symbolic Name |
| 18 | 0 | 121000 | N | 21:00:00:e0:8b:05:32:ec | 20:00:00:e0:8b:05:32:ec | NULL |
| 18 | 3 | 121300 | N | 21:00:00:e0:8b:06:dd:d4 | 20:00:00:e0:8b:06:dd:d4 | NULL |

Open a "Zone Admin" Window to identify the "WWN device ID" of your Fibre Channel device with a friendly name that will be easier to use in your SAN zoning configuration. ( create a Alias )

**Alias**

Alias Name: rafs04    [Create Alias] [Delete Alias] [Rename Alias]

Member Selection List — rafs04 Members
- SwitchPorts
- WWNs

rafs04 Members:
- 10:00:00:00:c9:23:a0:3c
- 20:00:00:00:c9:23:a0:3c

Open a "Zone Admin" Window to create a zone which allows access between the proper devices, using your preconfigured device aliases.

**Zone**

Zone Name: Rafs04    [Create Zone] [Delete Zone] [Rename Zone]

Member Selection List:
- SwitchPorts
- WWNs
- Aliases
- QuickLoops

Rafs04 Members:
- SwitchPorts
- WWNs
- Aliases
  - mti08
  - rafs04
- QuickLoops

Connectivity between the servers and disks has been established through the switches. Map the disk storage to the desired server using a utility provided by the manufacturer which assigns LUNS to servers using the WWN of the server's Fibre Channel Adapter.

| Host World Wide Name | SD #00 | SD #01 | SD #02 |
|---|---|---|---|
| Drive to LUN Mapping | *LUN0 | *LUN1 | *LUN2 |
| Allow *ANY* host access (currently defined hosts and any other host) | ☐ | ☐ | ☐ |
| 20-00-00-E0-8B-02-E5-CF Rmine001L | ☐ | ☑ | ☐ |
| 20-00-00-E0-8B-02-E3-CF Rmine001U | ☐ | ☑ | ☐ |
| Select *ALL* hosts defined above | ○ | ○ | ○ |

**Use of Bindings**

The following information illustrates the settings of the Fibre Channel HBA driver for use of bindings on the FC adapter card. This allows you to make the target device number static on all servers for the WWN of the device. This allows you to accurately track use of storage across your facility.

*Settings in qla2200.conf file:*

# MTI10
hba0-SCSI-target-id-220-fibre-channel-name= "21000cc001000939";
hba0-SCSI-target-id-221-fibre-channel-name= "23000cc001000939";
hba1-SCSI-target-id-220-fibre-channel-name= "21000cc001000939";
hba1-SCSI-target-id-221-fibre-channel-name= "23000cc001000939";

*Results of persistent binding configuration demonstrated using the "format" command:*

6. c4t220d0 <MTI-S200-7770 cyl 52066 alt 2 hd 256 sec 128>/pci@4,2000/scsi@1/sd@dc,0
7. c4t220d2 <MTI-S200-7770 cyl 52066 alt 2 hd 256 sec 128>/pci@4,2000/scsi@1/sd@dc,2
9. c4t221d1 <MTI-S200-7770 cyl 52066 alt 2 hd 256 sec 128>/pci@4,2000/scsi@1/sd@dd,1

**Multipathing**

When connecting storage to servers using switches and multiple HBAs in the server, there will be multiple images of each storage device presented to the server's operating system. This is desirable as each device presented is in reality another path between the server and its storage and provides higher availability to the storage. This configuration requires special software to handle the multiple paths to the storage without causing controller confusion or filesystem corruption. In our case Veritas Volume Manager is used to handle the mulipathing. Below, the output of a query on disk c2t185d1s2 shows the true 4 paths which are used to access it.

```
numpaths:  4
c2t185d1s2     state=enabled
c2t186d1s2     state=enabled
c3t185d1s2     state=enabled
c3t186d1s2     state=enabled
```

In the given example "c2" and "c3" are the 2 HBAs of the server, "t185" and "t186" are the 2 disk controllers of the RAID system, "d1" is the LUN id built on the RAID system and "s2" is the Solaris disk partition which represents the whole disk.

---

## Veritas Software:

Because it is an integral part of our disk management process, I will attempt to describe some of the basic concepts.

### Benefits:

. Physical Disk is unimportant
. Resize/Reconfigure Volumes during Realtime
. Mirroring and Moving data is transparent
. Provides organization and portability of data
. Multi-Pathing for High Availability

### Shortcomings of mounting common disks:

. Disk is known by O/S as a fixed name dependent upon hardware path.
. Disk can be divided into partitions which are permanent as long as data is on them.

### Integrating Veritas Volume Management:

. Once a disk is brought under volume manager control it becomes a Veritas resource and can be manipulated in many ways.

### Explaining Veritas Subdisks:

Rather than mounting a partition of a disk, Veritas creates a public partition encompassing all of the disk and allows you to use all or some of the storage on the disk as a virtual partition called a "subdisk"

*Once you're using a subdisk to mount a filesystem you can dynamically resize your partition.*

### Explaining Veritas Plexes:

Though your filesystems may often be smaller than your disk, sometime it is greater than the size of one disk. This necessitates the additional concept of a "plex".

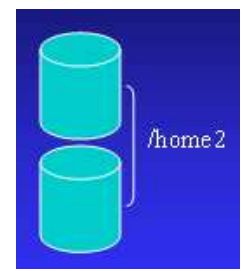

To have a mountable object that is larger than any single device, you must be able to glue storage together in either a *concatenated* or *striped* method. Concatenated disks write in one stream, filling in empty space as it is found on the multiple pieces. Striped plexes write to similar size subdisks concurrently, increasing performance with multiple I/O streams.

**Concatenated Plex**

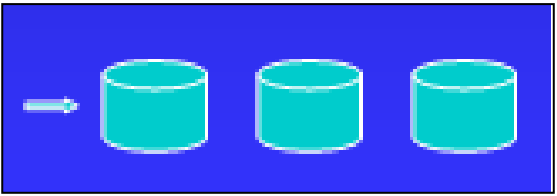

**Striped Plex**

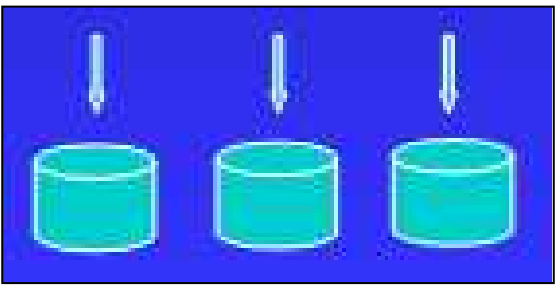

**Explaining Veritas Volumes:**

Sometimes you want multiple copies of your data (mirroring) This introduces the concept of a "volume". Since a plex has all the pieces which make up one copy of your data, a volume is made of one or one or more plexes which each have a copy of your data.

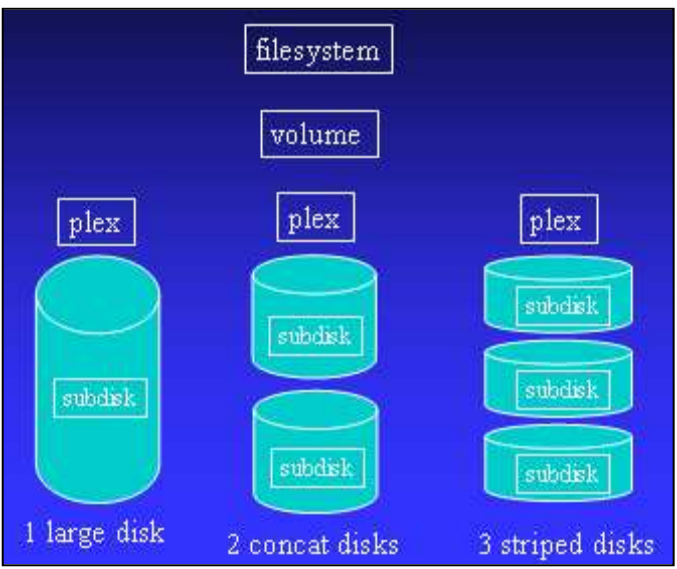

**Veritas Disk Groups:**

A disk group is the unit of **Portability** for Veritas Volume Manager. A group of disks can be moved between machines and irregardless of O/S identification of hardware, Veritas will put all of the pieces together. Below is an example of Disk Group components.

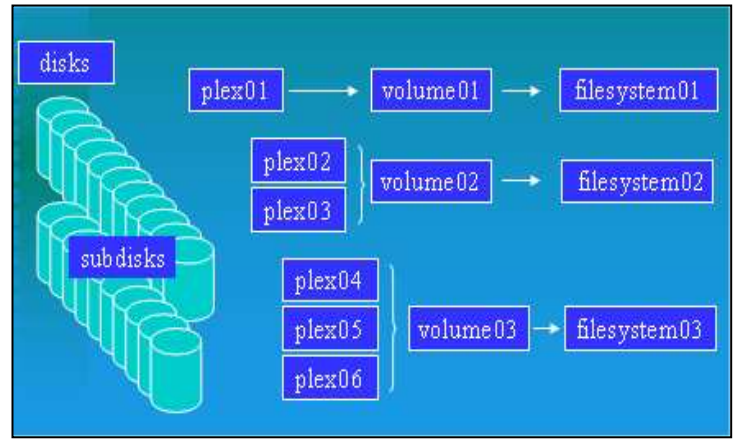